\documentclass[cits]{PoS}

\title{CNO Abundances of BA-Type Supergiants}

\ShortTitle{CNO Abundances of BA-Type Supergiants}

\author{\speaker{M.~Firnstein} and N.~Przybilla\\
        Dr.~Remeis-Sternwarte, Bamberg\\
        E-mail: \email{firnstein@sternwarte.uni-erlangen.de},
	\email{przybilla@sternwarte.uni-erlangen.de}}

\abstract{Massive BA-type supergiants are among the visually brightest
stars in galaxies with active star formation. As such they are versatile tools
for studies of stellar and galactochemical evolution. Moreover, they can act as
distance indicators for the calibration of the cosmological distance scale.
In the present work abundance patterns of the 
light elements helium, carbon, nitrogen and oxygen
are investigated in several Galactic BA-type supergiants in the mass range 
between 8 and 18 $M_{\odot}$. Based on high-resolution and high-S/N
Echelle spectra obtained with FOCES on the Calar Alto 2.2m telescope, model
atmosphere analyses are performed using state-of-the-art non-LTE spectrum
synthesis. Stellar parameters and chemical abundances are determined with
high accuracy. This gives tight observational constraints on the evolutionary
status of the stars. Objects evolving from the main sequence to the red
supergiant stage and those on a blue loop can be distinguished by their
mixing signature (pure rotational vs. first dredge-up). The most sensitive
tracer of nuclear processed matter, the N/C ratio, indicates a higher
mixing efficiency than predicted by current evolution models of rotating
stars with mass-loss.}

\FullConference{International Symposium on Nuclear Astrophysics --- Nuclei in the Cosmos --- IX\\
		 June 25-30 2006\\
		 CERN, Geneva, Switzerland}

\begin{document}

\section{CNO-Abundances and the Evolution of Massive Stars}
BA-type supergiants are among the visually brightest massive stars, allowing for a mapping of 
the spatial distribution of elemental abundances throughout galaxies and therefore a 
determination of abundance gradients. They can be versatile indicators for studies of 
stellar and galactic evolution as well as in the cosmological context. 
Comprehensive quantitative understanding of BA-type supergiants in the Milky Way is an 
essential step on the way to use this potential. 

An introduction to current stellar evolution models of massive stars can be found in \cite{Maed2000} or \cite{Heger}. 
Stars on the main sequence (MS) -- during the longest part of their life -- burn hydrogen to helium in
their cores. For massive stars with their high core temperature and pressure the most
efficient process for this turns out to be the CNO-bi-cycle.  While the actual elements carbon, nitrogen and
oxygen only function as catalysts, meaning no net production or loss occurs
during a cycle, their relative numbers change over the period of hydrogen
burning. The reaction  ${}^{14}\!$N$ (p,\gamma){}^{15}\!$O is a bottleneck, because it 
is the slowest by far. The abundance of the stable nitrogen isotope increases while the
abundances of the stable carbon and oxygen isotopes decrease, with the most abundant 
oxygen isotope ${}^{16}\!$O being transformed more slowly in the minor NO-branch of the cycle.

Stellar spectra trace abundances only of the photospheric layers. 
In order for processed matter to be recognisable, it has to travel the long way from the
centre of the star to the outer layers. Massive stars on the main sequence possess no
convective envelope like the Sun, ruling out convection as transport mechanism. 
However, slower mixing mechanisms exist, like rotation-induced meridional circulation or 
shear mixing (caused by differential rotation). Their efficiency increases with both rotational
velocity and mass of the star and can also depend on the magnetic field strength.

BA-type supergiants have already left the main sequence, their progenitors being OB-stars.
Their envelope is still in radiative equilibrium, so that no major changes in the abundance patterns are
expected during the short post-MS phase, should they have evolved directly
from the MS. For less-massive
supergiants another scenario is possible, the so called blue loop. Such objects have already 
reached the red supergiant stage, but eventually evolve back into a blue supergiant. During the
red supergiant phase, their envelope has been fully convective and intense
mixing has taken place (first dredge-up). This strengthens the signature of CNO-processed material
in the photosphere considerably, allowing to distinguish them from the other class of
supergiants.\\

\section{Observations and Quantitative Spectral Analysis}

Quantitative spectroscopy with high precision
is required to derive information on the evolutionary status of BA-type
supergiants. On the observational side high-resolution and high-S/N spectra have been 
obtained using the Echelle spectrograph FOCES at the Calar Alto 2.2m telescope. Therefore objects in the solar neighbourhood -- preferably associated to open clusters -- of spectral types B5-A3 and luminosity classes Ib to Ia have been chosen, in order to cover this part of the HRD as completely as possible. To extract 
the abundance information from the spectra, they are compared to
synthetic spectra using the fit routine FITPROF (see \cite{Napi} for a description). 
Previous quantitative analyses of Galactic BA-SGs are mostly based on standard LTE techniques, which provide only limited accuracy. Recently, we introduced a NLTE spectrum synthesis technique that allows for high-precision analyses of BA-SGs (\cite{Norb}). NLTE line-formation computations are performed with DETAIL and
SURFACE on the basis of hydrostatic line-blanketed
LTE model atmospheres (ATLAS9) in plane-parallel geometry, see Fig.~1 for an
overview. This method takes new state-of-the-art model atoms for H , 
He\,{\sc i}, C\,{\sc i/ii}, N\,{\sc i/ii}, O\,{\sc i} and Mg\,{\sc i/ii} into account, based
on accurate atomic data and modern line-broadening theories. This allows for a highly accurate determination of metal abundances with low statistical scatter (10-20 \% on the $1\sigma$-level) and leads to highly consistent abundances for individual stars (\cite{Flor}) as well as for a larger sample of objects (this work).

\begin{figure}
\centering
\includegraphics[width=.7\textwidth]{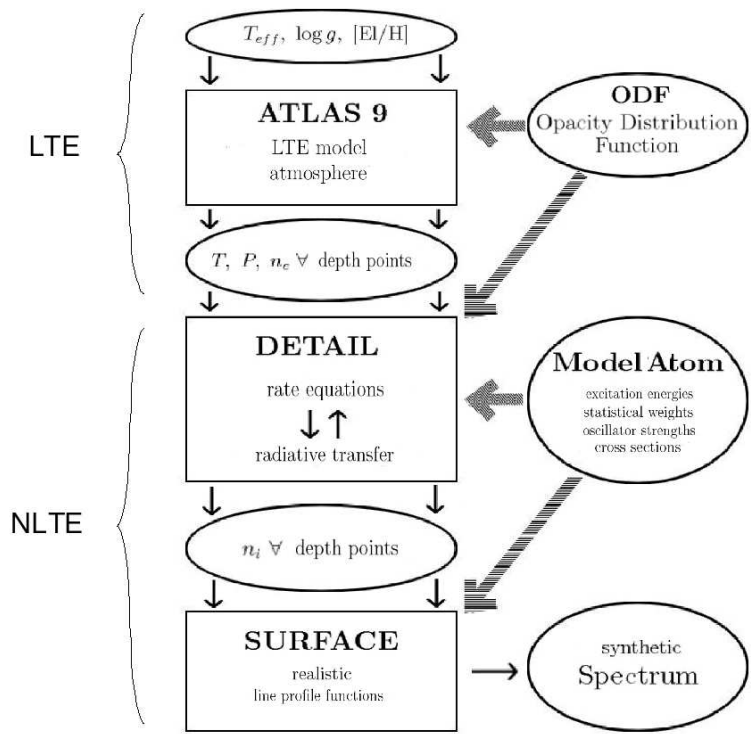}
\caption{Overview of the model calculation process}
\label{atsurf}
\end{figure}

The stellar parameters are determined via line-profile fits to the Stark-broadened 
hydrogen Balmer and He\,{\sc i} lines, as well as ionization equilibria such as
Mg\,{\sc i/ii} for cooler stars and N\,{\sc i/ii} for hotter stars of the sample. The method
is described by \cite{Norb}.

While neither of the criteria alone allows the unambiguous determination of
effective temperature $T_{\rm{eff}}$ and surface gravity $\log g$, a combination
of them does. 
Problems may arise for H${\alpha}$, as stellar wind effects are neglected in
our modelling, the effects being strongest in the most massive and therefore most luminous supergiants.

High accuracy in the parameter determination is required in order to derive 
chemical abundances with high precision. Not only abundance, $T_{\rm{eff}}$ and $\log g$ influence the line
strength, but also the microturbulence and the helium content. After determining
all these parameters by iteration we can achieve accuracies in chemical
abundances of better than 0.1\,dex, depending on the number of available lines and the
quality of the spectrum.

\section{Results and Conclusions}

\begin{figure*}
\centering
\includegraphics[height=.70\textwidth,angle=-90]{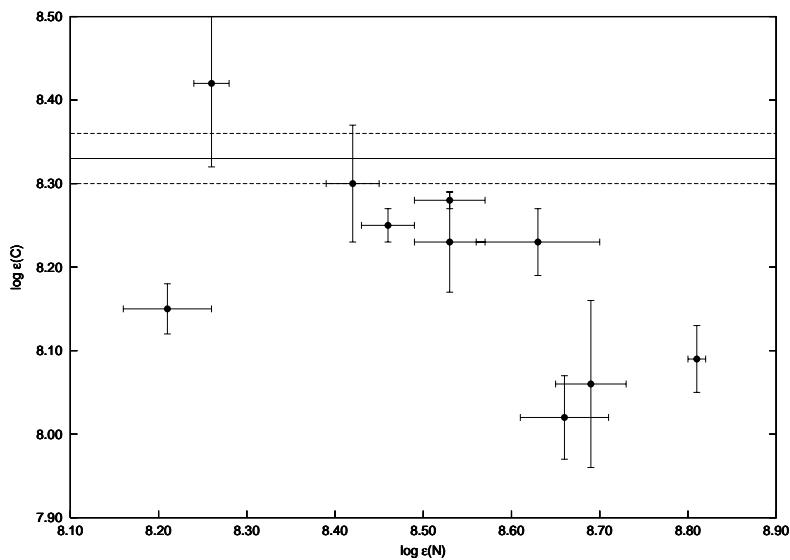}
\caption{Carbon abundances as a function of the nitrogen abundance for the 10
stars of the sample, where $\log\varepsilon({\rm X}) =
\log(N_{\rm{X}}/N_{\rm{H}})+12$.  
They are compared to the mean value $8.33 \pm 0.03$ obtained by \cite{Nieva} from carbon
abundances in 6 B-dwarfs and
giants}
\label{CN}
\end{figure*}

The abundance data for the 10 analysed BA-Type supergiants show some interesting
results. While magnesium and oxygen abundances show only a small scatter, the expected trends 
for carbon and in particular nitrogen become
apparent (see Fig.~\ref{CN}).

\begin{figure*}
\centering
\includegraphics[width=.95\textwidth]{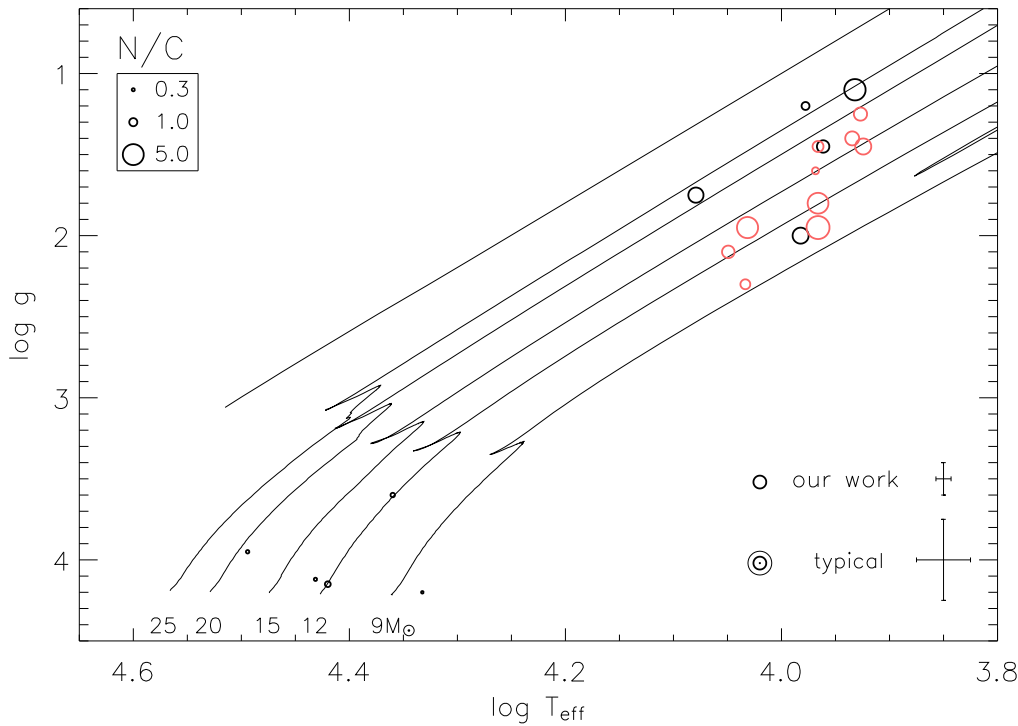}
\caption{Data for objects analysed in this work (coloured circles) are
compared to evolution tracks from \cite{Maed3} for rotating stars($v_{\rm{rot}}=300$km/s). The models start with a N/C ratio
of 0.3 on the main sequence and predict ratios of 0.7 to 1.7 for
BA-type supergiants in the mass range of 9 to 25 $M_{\odot}$ evolving redwards and 1.6 to
2.6 for stars of 9 to 12 $M_{\odot}$ on a blue loop, while our findings for 10 BA-type supergiants show N/C ratios between 0.81 and 6.16. Also displayed are additional results for BA-type
supergiants from \cite{Norb} and B-dwarfs and giants from \cite{Nieva}, all analysed in a homogenous way.}
\label{evol}
\end{figure*}

In combination with a precise determination of the stellar parameters the
abundance analysis allows comparisons with stellar evolution models. Apparently
the N/C ratios are generally larger than predicted by theory, and significantly
larger than constrained from observation before (\cite{Venn1},\cite{Venn2}: we have 
four objects in common with those studies). More recent
stellar evolution computations accounting for the interplay of rotation and
magnetic fields, as described by \cite{Maed5}, promise to resolve these discrepancies
as they show a much higher efficiency for chemical mixing. A possibly even better solution provide newer measurements of the ${}^{14}\!$N$ (p,\gamma){}^{15}\!$O cross section (\cite{Reaction}), that reduce previous estimates of the reaction rate by a factor of 2. High
N/C ratios also particularly occur in regions where blue loops are predicted, implying
that these objects have undergone the first dredge-up (see Fig.~\ref{evol}). However, 
the blue loops may need
to extend further to the blue than predicted by the current models.

\end{document}